\documentclass[prb,twocolumn]{revtex4}
\usepackage[latin1]{inputenc}
\usepackage{hyperref}
\usepackage{amsmath}
\usepackage{graphicx}
\usepackage{graphics}
\usepackage{amsfonts}
\usepackage{amssymb}
\usepackage{natbib}
\usepackage{subfigure}
\begin{document}
\title{The Mixed State of a $\pi$-Striped Superconductor}
\date{\today}
\author{M. Zelli}
\email{zellim@mcmaster.ca}
\affiliation{Department of Physics and Astronomy, McMaster University, Hamilton, Canada}
\author{Catherine Kallin}
\affiliation{Department of Physics and Astronomy, McMaster University, Hamilton, Canada}
\author{A. John Berlinsky}
\affiliation{Department of Physics and Astronomy, McMaster University, Hamilton, Canada}
\begin{abstract}
A model of an anti-phase modulated d-wave superconductor has
been proposed to describe the decoupling between Cu-O
planes in $1/8$ doped La$_{2-x}$Ba$_{x}$CuO$_{4}$. Unlike
a uniform d-wave superconductor, this model exhibits an
extended Fermi surface. Within Bogoliubov-de Gennes theory, we study the mixed state of  this
model and compare it to the case of a uniform d-wave superconductor.
We find a periodic structure of the low-energy density
of states, with a period that is proportional to $B$, corresponding to Landau levels that are a coherent mixture of particles and holes.  These results are also discussed in the context of experiments which observe quantum oscillations in the cuprates, and are compared to those for models in which the Fermi surface is reconstructed due to translational symmetry breaking in the non-superconducting state and to a model of a Fermi-arc metal. 
\end{abstract}
\maketitle
\section{Introduction}
High temperature superconductivity in the underdoped cuprates condenses from a so-called pseudogap phase whose properties are distinctly different from those of a conventional metal.\cite{pseudo,timusk} Below a relatively high temperature, $T^*$, a gap which may or may not be connected to superconductivity starts to develop in the excitation spectrum and affects the temperature dependence of all transport processes. At lower temperatures, above but closer to $T_c$, superconducting fluctuations in the form of a disordered vortex-antivortex liquid grow up until long-range d-wave superconducting order appears at $T_c$.\cite{antiv}

The nature of the pseudogap phase has been the subject of much study and debate. One can characterize the behavior in this phase as ``non-Fermi-liquid-like", which typically means that the sharp fermionic excitations of a Fermi liquid are broadened even close to the Fermi surface (FS), although the situation in the pseudogap is somewhat more complicated. ARPES observes ``Fermi arcs" -- sections of 2D FS which appear to terminate at gaps and which become shorter, possibly tending toward nodal points, as $T$ is lowered.\cite{arc,fermiarc} No sharp quasiparticles are observed near the anti-nodal points at $(\pm\pi,0)$ and $(0,\pm\pi)$.

The origin of this non-Fermi-liquid-like behavior is hotly disputed. Some studies connect it to resonating valence bonds or preformed pairs,\cite{rvb} while others associate it with exotic forms of fluctuating or static spatial order such as charge or spin density waves\cite{densitywave,cdw} or singlet or triplet D-density waves.\cite{chak}

This unsatisfactory state of understanding was compounded in 2007 by the observation of quantum oscillations, first in the Hall resistivity and shortly afterward in Shubnikov-de Haas and in de Haas-van Alphen measurements, below $T_c$ at fields that are comparable to, although typically smaller than, $H_{c2}$.\cite{doiron,jaudet,riggs,bangura,yelland,audouard,sebastian,singleton,rourke} Since such fields are still weak relative to Fermi energy scales, one might expect that, once superconductivity is destroyed by a large magnetic field, the underlying resistive state would be the same pseudogap state as exists above $T_c$, that is, a non-Fermi liquid. However quantum oscillations are normally associated with sharp, {\em closed} Fermi surfaces of a Fermi liquid. Furthermore, it is not straightforward to connect the FS areas determined by quantum oscillations with the Fermi arc observed in ARPES.\cite{onsager} However it has been noted that commensurate static translational symmetry breaking, due to charge or spin density waves, could reorganize the large hole FS of the undistorted lattice into a number of smaller hole and electron pockets and that the small electron pockets could account for the quantum oscillations, while sections of the larger hole pockets coincide with the Fermi arc.\cite{millis,harrison} The explanation for the arcs is then that the spectral weight due to the periodic perturbation of the charge-density wave (CDW) or spin-density wave (SDW) is large on the arcs that are observed by ARPES and small on the remainder of the FS hole pockets. Such a result is consistent with a simple picture of zone folding due to a weak periodic superlattice potential.

In this paper, we consider a variation of this picture in which the periodic superlattice arises from a modulation of the d-wave superconducting gap function.\cite{berg,stripe,shirit} Such a modulation has been invoked to explain an interesting phenomenon called the $1/8$ anomaly which is observed in some of the lanthanum cuprates.\cite{kivelson} Most lanthanum cuprates exhibit singular behavior in the doping dependence of various low-temperature properties around $1/8$ doping, which is known to coincide with a charge stripe structure with the periodicity of 4 lattice constants. The superconducting condensate for the proposed model occurs at a nonzero wave vector, corresponding to a period twice that of a charge stripe structure.\cite{berg,stripe} Furthermore, the stacking arrangement assumed for this model results in a zero Josephson coupling between nearest cuprate layers which explains the apparent dynamical decoupling of cuprate layers observed in transport measurements of $1/8$ doped La$_{2-x}$Ba$_{x}$CuO$_{4}$. \cite{qLi} In this paper, we study such a modulated d-wave gap in the presence of large magnetic fields using lattice Bogoliubov-de Gennes (BdG)  theory\cite{degennes} to determine if quantum oscillations, associated with Landau level formation, occur.

The remainder of this paper is organized as follows. In Sec. 2, we describe the model for a $\pi$-striped superconductor. This section shows the density of states (DOS), the spectral functions and the FS in zero field for different amplitude gaps for this model. In Sec. 3, we establish the generation of Landau levels in the DOS by a magnetic field and the effect of doping and of the gap amplitude on the Landau level spectra. In Sec. 4, the specific heat is calculated to make some connections to the experiments.  Finally, in Sec. 5, we discuss the results and compare them to other models and to relevant experiments.
\section{The Model and Method}
The two-dimensional tight-binding model for a $\pi$-striped superconductor is described by the following mean-field Hamiltonian\cite{shirit}
\begin{align}
H=H_{0}+\sum_{x,y} \Delta \{ \cos(q_{x}x)[c^{\dagger}_{x,y\uparrow}c^{\dagger}_{x+1,y\downarrow}-c^{\dagger}_{x,y\downarrow}c^{\dagger}_{x+1,y\uparrow}] \label{eq:hamil} \\
-\cos(q_{x}(x-1/2))[c^{\dagger}_{x,y\uparrow}c^{\dagger}_{x,y+1\downarrow}-c^{\dagger}_{x,y\downarrow}c^{\dagger}_{x,y+1\uparrow}]+H.C. \} \nonumber
\end{align}
where $c^{\dagger}_{x,y\sigma}$ creates an electron with spin $\sigma$ on site $(x,y)$. By setting $q_{x}=\pi/4$, the Hamiltonian describes a system with a d-wave-type order parameter that has a sinusoidal modulation with $8$-site periodicity in the $x$ direction. $H_{0}$, the kinetic part of the Hamiltonian, has the dispersion $\epsilon_{0}= -2t (\cos(k_{x})+\cos(k_{y}))-4t_{2}\cos(k_{x})\cos(k_{y})-\mu $ in $k$ space where $t$ and $t_{2}$ are the first and second nearest neighbor hopping terms. Due to the periodic modulation of the order parameter, the superconducting condensate occurs at a non-zero $q$, and a particle with wave vector $k$ is coupled to ones with wave vectors $-k\pm q_{x}$. We shall see that this property of a striped superconductor has crucial effects on its low-energy properties.

\begin{figure} [tbph] 
\includegraphics[scale=.35]{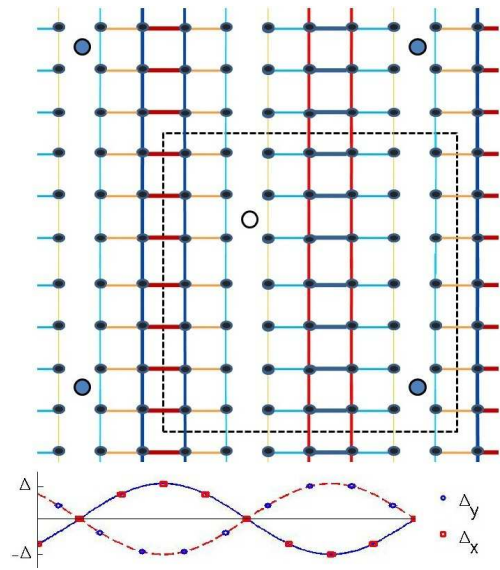}
\caption{Position dependence of the pairing gap for the bond-centered configuration using color coding on bonds. The circles in the middle of the plaquettes specify the positions of vortices for a $l=8$ magnetic field unit cell whose boundary is shown by the dashed line. In the singular gauge, the vortices at white (dark) circles are only seen by particles (holes). The lower part of the figure shows the varying gap amplitude as a function of $x$.} 
\label{config}
\end{figure}
\begin{figure} [tbph] 
\includegraphics[scale=.22]{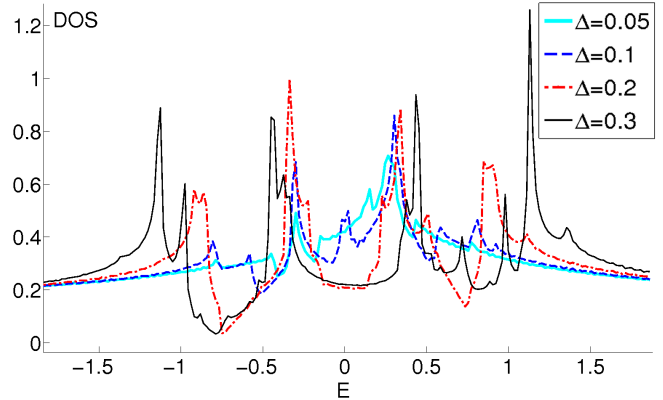}
\caption{DOS of a $\pi$-striped superconductor for various values of the pairing gap amplitude. The second nearest neighbor hopping for this DOS calculation is set to zero and the chemical potential $\mu$ is adjusted to yield $1/8$ doping here and in the following figures unless another value is explicitly stated. Note the finite DOS at zero energy and the complex structure which arises from band folding associated with the strength of the periodic interaction as discussed in the text.} 
\label{fig:dosall}
\end{figure}
There are two possible stable configurations for the order parameter of the $\pi$-striped superconducting model. One configuration is the site-centered configuration in which the node of the modulation lies on a site. The other one is the bond-centered configuration in which the node lies on a bond. The calculations in this work are done for the latter configuration which is shown in Fig. \ref{config}. However, the qualitative behavior of the system in the presence of a magnetic field is similar for the site-centered case.\par
\begin{figure} [tbph] 
\includegraphics[scale=.57]{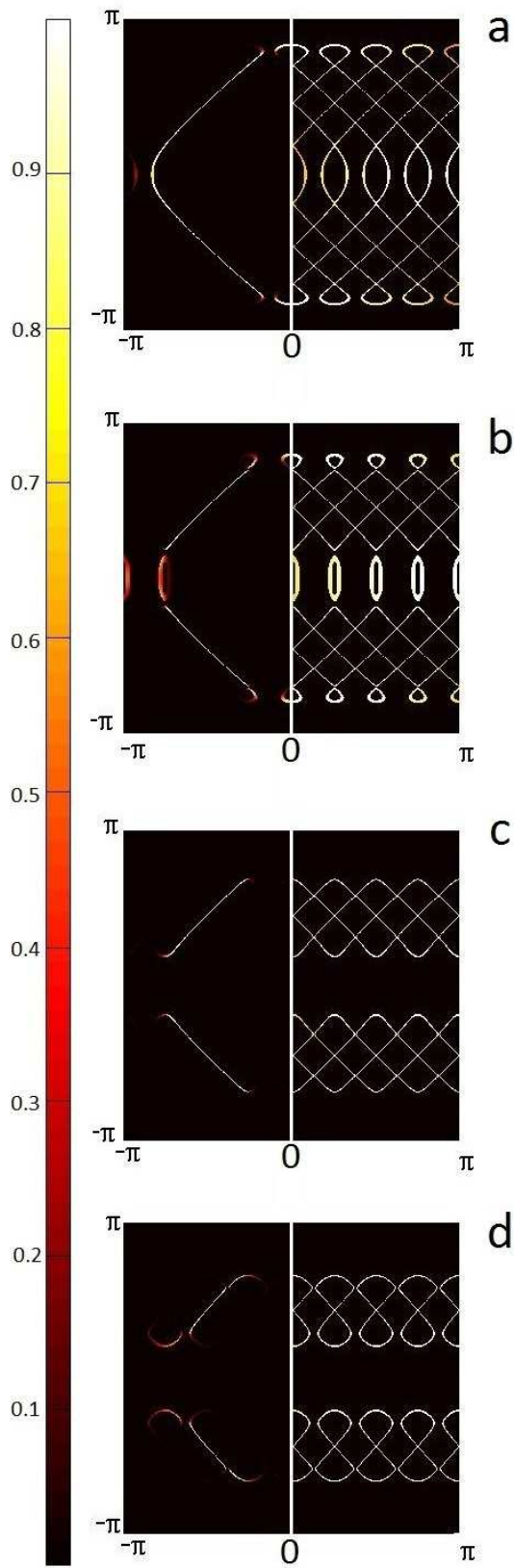}
\caption{ The spectral weight (left) and FS (right) for four values of the pairing gap $\Delta$ a) $0.05$, b) $0.1$, c) $0.2$ and d) $0.4$ in half of the extended BZ. The colorbar applies only to the spectral weight.} 
\label{spall}
\end{figure}
We solve the model by diagonalizing the BdG Hamiltonian.\cite{degennes} The local density of states is defined as 
\begin{align}
D(i,E)=2\sum_{n=1}^{Nl^2}{[|u_{n}(i)|^{2}\delta(E-E_{n})+|v_{n}(i)|^{2}\delta(E+E_{n})]}
\end{align}
where $l^{2}$ is the number of sites in one magnetic unit cell, $N$ is the number of magnetic unit cells and ($u_{n}(i)$,$v_{n}(i)$) is the position eigenstate of the $n$-th positive-energy state. Due to translational symmetry, the local density of the system is the same for all unit cells. The spectral weight function in the extended Brillouin zone (BZ) is defined as 
\begin{align}
A(k,E)=\frac{2}{Nl^{2}}\sum_{n=1}^{Nl^{2}}{[|u_{n}(k)|^{2}\delta(E-E_{n})+|v_{n}(k)|^{2}\delta(E+E_{n})]}
\end{align}
where $u_{n}(k)$ ($v_{n}(k)$) is the Fourier transform of $u_{n}(i)$ ($v_{n}(i)$). In the reduced BZ scheme, one sums $A(k,E)$ over the eight coupled $k$ in the extended BZ that can be folded back to one point in the reduced BZ. The DOS as a function of energy can be obtained from the position average of the local density of states or the wave vector average of the spectral weight function.

The DOS of a homogeneous d-wave superconductor vanishes linearly at low-energy. In contrast, a $\pi$-striped superconductor has a non-zero DOS at zero energy. \cite{shirit} The low-energy dependence of the DOS for various values of the pairing gap amplitude, $\Delta$, is shown in Fig. \ref{fig:dosall}. For simplicity, the second nearest neighbor hopping in the kinetic part of the Hamiltonian has been set to zero and all the energy quantities are written in units of the nearest neighbor hopping $t$ which is set to 1. The chemical potential, $\mu$, is adjusted to yield $1/8$ doping unless another value is explicitly stated.

It is useful to compare and contrast the $\Delta$ dependence of the DOS shown in Fig. \ref{fig:dosall} to that of the one-electron spectral weight. For small $\Delta$, small gaps open in the unperturbed FS segments that can be connected by $q_{x}$ as shown in Fig. \ref{spall}(a). Consequently the DOS at zero energy does not change significantly with respect to the unperturbed case. For intermediate values of $\Delta$, in the approximate range ($0.07\lesssim \Delta \lesssim 0.13$), the pairing gap has become strong enough to create two sets of closed loops which can be seen in Fig. \ref{spall}(b) for $\Delta=0.1$. The Fermi velocity associated with these loops is small and consequently they contribute considerably to the DOS at zero energy. This is why there is a peak in the DOS for $\Delta=0.1$. For larger values of $\Delta$, in the approximate range ($0.14\lesssim \Delta \lesssim 0.25$), the loops are gapped out and the peak disappears. In this range, the spectral weight exhibits Fermi arcs with two small gaps. For even larger $\Delta$, the gaps within the Fermi-arc-shaped spectral weight become larger and the shape of the FS in the repeated BZ scheme appears as figure-$8$-shaped loops as shown in Fig. \ref{spall}(d) for $\Delta=0.4$.

\section{Results in a Magnetic field}\label{section:dos}
The above results are all for zero magnetic field. A magnetic field is incorporated into the model using the Franz-Tesanovic singular gauge transformation.\cite{fttrans,dwaveft} In this approach, the magnetic unit cell has linear size $l$, where $l$ is measured in units of the lattice constant. Each unit cell has two vortices; one is seen only by particles and the other seen only by holes. We position the vortices at the nodes of the order parameter, as shown in Fig. \ref{config} for the case of $l=8$. We take $l$ to be an integer multiple of $8$ which is the period of the order parameter. The magnetic field and $l$ are related by $B=\phi_{0}/l^{2}$ where $\phi_{0}$ is the flux quantum. For example, taking the lattice constant $a=3.8 \AA$, $l=32$ corresponds to $B=28$ T. Hereafter, we express the magnetic field in terms of $l$.

\begin{figure} [tbph] 
\includegraphics[scale=.33]{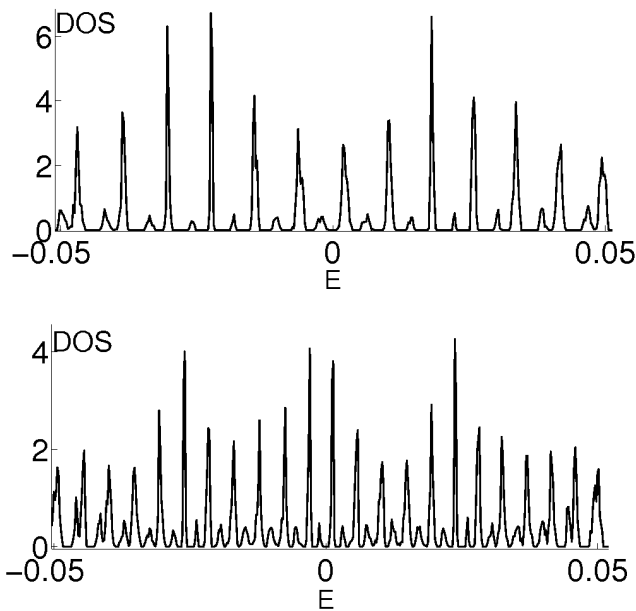}
\caption{Low-energy DOS of a $\pi$-striped superconductor with $\Delta=0.01t$ and $\mu=-0.226$ in the presence of magnetic fields of $l=24$ (top) and $l=32$ (bottom).} 
\label{dsmalldos}
\end{figure}
In this section, we investigate how the DOS structure of the model changes as a function of the pairing amplitude $\Delta$ in the presence of a magnetic field. First, we consider small values of $\Delta$ where one can expect to understand the effect of the interaction based on a simple perturbative picture. For $\Delta=0.01$, the spectral weight in the absence of a magnetic field exhibits only small gaps at four points of the unperturbed FS. It is similar to Fig. \ref{spall}(a) except that the gaps are smaller.

The low-energy DOS structures for $\Delta=0.01$ and two magnetic fields of $l=32$ and $l=24$ are shown in Fig. \ref{dsmalldos}. The most striking feature of this figure is the appearance of Landau levels that are equally spaced in energy with the spacing proportional to $B$. Furthermore, the presence of a small perturbative interaction, $\Delta$, causes the low-energy Landau levels to be slightly broadened and also partially reflected to the other side of the Fermi energy due to particle-hole scattering. In fact, each Landau level for $\Delta=0$ is split into two peaks with the second peak having much smaller weight for small $\Delta$, as seen in Fig. \ref{dsmalldos}. The sum of the number of states in these two peaks equals the degeneracy of a Landau level.

From a semiclassical point of view, particles can keep undergoing Larmor precession by tunnelling through the gaps since the gaps are small for $\Delta=0.01$. This is the so-called magnetic breakdown phenomenon.\cite{shoenberg} A particle can also be Andreev scattered as a hole into a state of $-k\pm q$. This process explains the reflected part of each Landau level with smaller weight in Fig. \ref{dsmalldos}. This picture is motivated by the work of Pippard,\cite{pippard1962} who studied the cyclotron motion of nearly free electrons in the presence of a weak periodic potential that induces gaps in the Fermi surface. For this case, when the periodic potential is weak, electrons can tunnel through the gaps, following the unperturbed FS trajectory, or they may be Bragg scattered onto a different cyclotron orbit leading to broadening. The main difference between Pippard's model and the $\pi$-striped superconducting model is that the superconducting periodic potential scatters electrons with wave vector k into holes with wave vector $-k\pm q$ and vice versa. Thus electrons either tunnel through gaps induced by the periodic potential or scatter into hole states. Note that magnetic breakdown occurs even if the magnitude of the gap in the FS is larger than $\hbar \omega_{c}$.\cite{shoenberg}

For intermediate values of $\Delta$ ($0.07\lesssim \Delta \lesssim 0.13$), for which the FS has well-separated segments (see Fig. \ref{spall}(b)), we do not observe clearly defined Landau levels. This may be the result of broadening and the close spacing of Landau levels due to the large density of states. Furthermore, multiple Fermi surfaces may each give rise to their own sets of Landau levels which are unresolved.

\begin{figure} [tbph] 
\includegraphics[scale=.22]{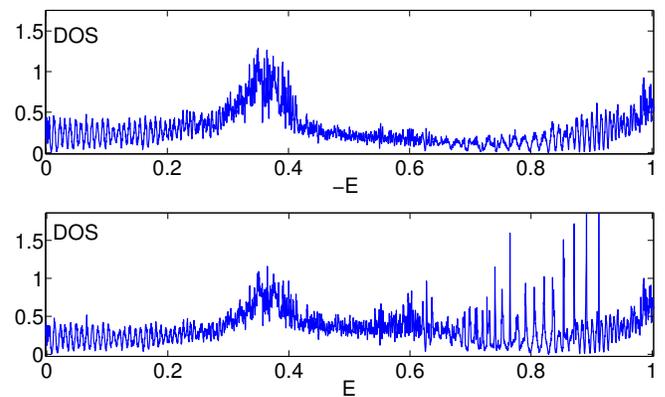}
\caption{DOS for $\Delta=0.25$ and magnetic field of $l=32$ shown as a function of positive and negative energies separately. The band structure spans energies from $-4-\mu$ to $4-\mu$. However, the DOS  is only shown in the $-1<E<1$ range.} 
\label{dos32}
\end{figure}
\begin{figure} [tbph] 
\includegraphics[scale=.22]{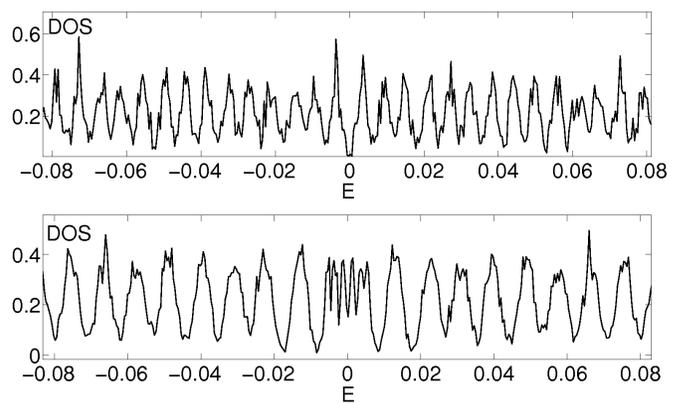}
\caption{Low-energy DOS for $\Delta=0.25$ and magnetic fields of $l=40$ (top) and $l=32$ (bottom).} 
\label{32and40}
\end{figure}
For larger values of $\Delta$ ($0.14\lesssim \Delta \lesssim 0.3$), the shape of the FS is simpler. In this range, the spectral function has significant weight on the parts of the FS that resemble the Fermi arcs observed in ARPES experiments. The DOS for $\Delta = 0.25$ and $l = 32$ is shown in Fig. 6 for positive and negative energies up to E = 1 separately. Remarkably, we again observe periodic behavior of the low-energy DOS as a function of $E$ with a spacing that varies linearly with B.  This is illustrated in Fig. \ref{32and40} for $\Delta = 0.25$ and two values of the magnetic field, $l = 40$ and $l = 32$. Note the splitting of each Landau level into a strong and weak peak seen for small $\Delta$, Fig. \ref{dsmalldos}, does not occur in this larger $\Delta$ range, where the original large FS is not accessible to the quasi-particles.

In Fig. \ref{32and40}, the DOS has a minimum, or possibly a very small gap, at $E=0$ for $l=40$. However, for $l=32$ it appears that the two Landau levels closest to $E=0$ are joined together and the DOS at $E=0$ has a nonzero value. In general, we find that, for $l=8m$ where $m$ is an integer, if $m$ is even, the DOS at $E=0$ is nonzero and if $m$ is odd, the DOS is zero at $E=0$. This is a commensurability effect that results in oscillation of the $DOS$ at $E=0$ as a function of $l$ or $1/\sqrt{B}$ and is discussed further in appendix \ref{a1}.
\begin{figure} [tbph] 
\includegraphics[scale=.195]{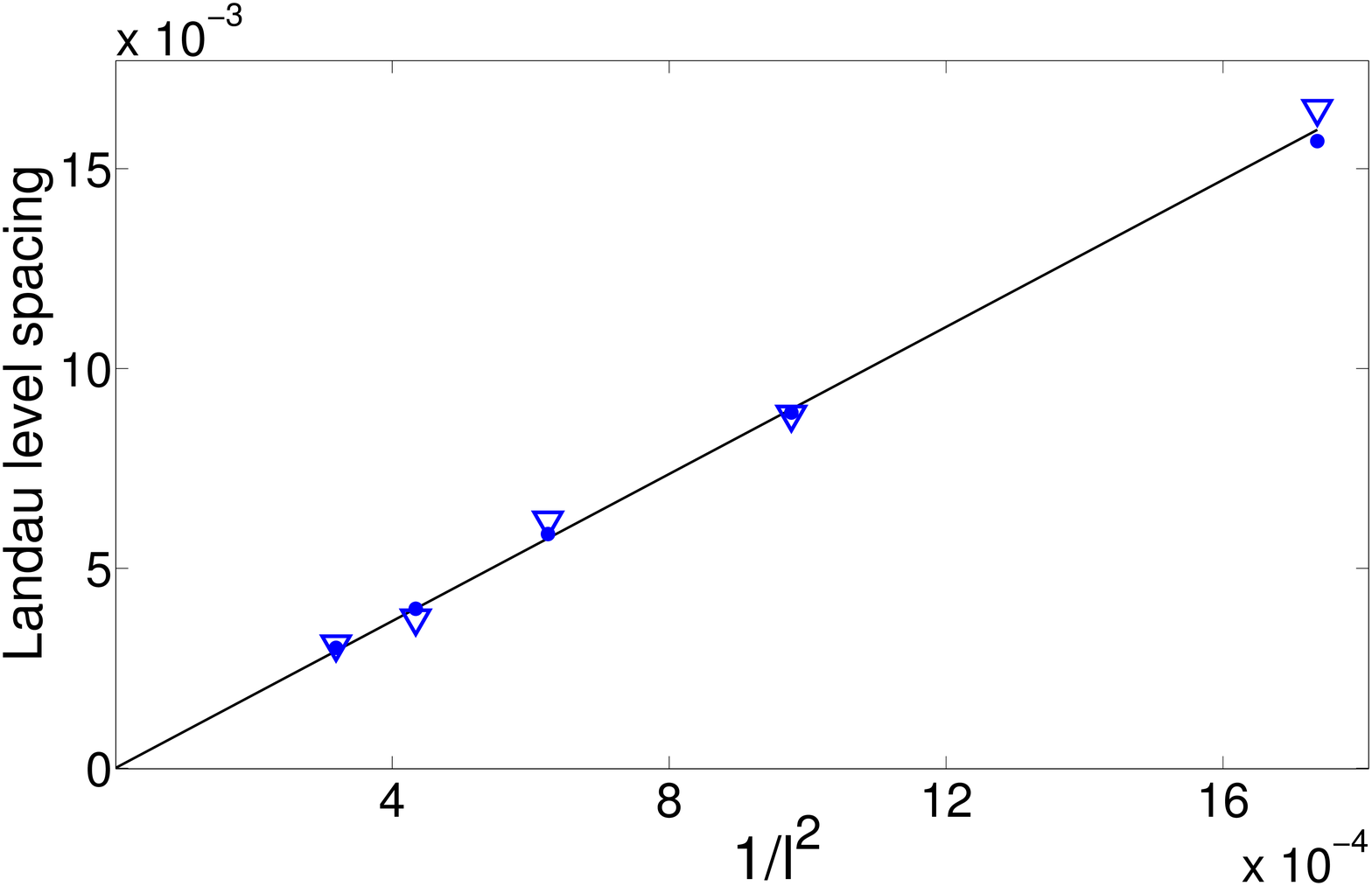}
\caption{Low energy Landau level spacing as a function of $1/l^{2}$ for $\Delta=0.25$ and $\mu=-0.3$. The spacing is defined as $E(N)/N$ where $E(N)$ is the minimum in the DOS between the $N$-th and $N+1$-th Landau levels and is shown for $N=2$ (triangle) and $N=10$ (circle). The line is a linear fit to the data that goes through the origin.} 
\label{TvsB}
\end{figure} 

We have calculated the spacing of the low energy Landau levels for a wide range of fields for $\Delta = 0.25$ and $\mu=-0.3$ as shown in Fig. \ref{TvsB}. The Landau level spacing can be defined as $E(N)/N$ where $E(N)$ is the minimum in the DOS between the $N$-th and $N+1$-th Landau levels and is essentially independent of $N$ provided $E(N)\lesssim 0.5\Delta$. Fig. \ref{TvsB} demonstrates the spacing as a function of $1/l^{2} \propto B$ for $N=2$ and $N=10$. The slope of the Landau level spacing versus $B$ is inversely proportional to the DOS at $E = 0$. By comparison, we find that the slope is about half as large and the DOS at $E = 0$ about twice as large for $\Delta = 0$.

\begin{figure} [tbph] 
\includegraphics[scale=.22]{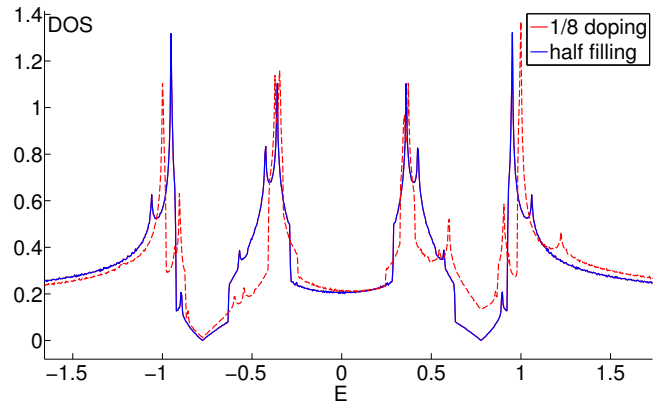}
\caption{The DOS structure in the absence of a magnetic field for $\Delta=0.25$ and two dopings. Note the asymmetry at low $E$ for $1/8$ doping.} 
\label{half}
\end{figure}

Furthermore, the number of states in each peak is nearly the same as that of a Landau level. In general, in the presence of a magnetic field, the $n$-th peak on the left of $E=0$ can have a degeneracy slightly different from a Landau level degeneracy. However, the $n$-th peak on the right compensates so that the number of states of the two peaks together is always twice that of a Landau level. This shows that the Landau levels are a coherent mixture of particles and holes together and the particle-hole scattering is playing a role in the formation of the Landau levels. The reason for the small difference of the number of states in each peak from the exact degeneracy of a Landau level is that the low-energy DOS in the absence of a magnetic field is asymmetric around $E=0$ except at half filling, as shown in Fig. \ref{half}. 

It is worth contrasting the behavior of the $\pi$-striped superconductor, Fig. \ref{32and40}, to the DOS structure of a homogeneous d-wave superconductor. For the latter at half-filling, Landau levels are formed in the low-energy DOS, but the Landau levels are not equally spaced and the spacing scales as $\sqrt{B}$ around $E=0$.\cite{dwaveft} This is a consequence of the nodal behavior at the Fermi energy. Therefore, quantum oscillations periodic in $1/B$ are not expected for a d-wave superconductor. For the remainder of this paper we refer to each peak of the type shown in Fig. \ref{32and40} (that is, equally spaced with a spacing proportional to $B$) as a Landau level. The fact that there is only one set of Landau levels and the number of states in each peak is equal to that of a Landau level indicates that all parts of the FS participate in the formation of low-energy Landau levels.
As noted above, the case of large values of $\Delta$, Fig. \ref{32and40}, is different from the case of very small values of $\Delta$, Fig. \ref{dsmalldos}, in which the sum of the degeneracy of the two peaks equals the degeneracy of only one Landau level. 
\begin{figure} [tbph] 
\includegraphics[scale=.21]{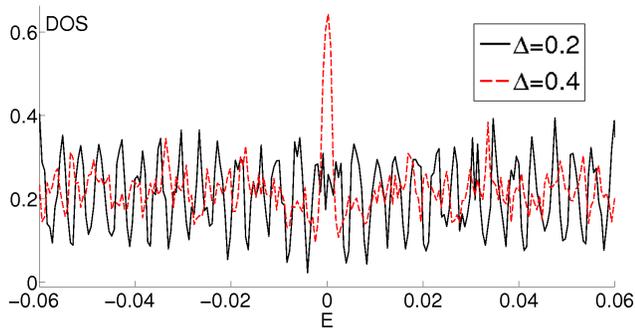}
\caption{Low-energy DOS for $l=48$ and $\Delta=0.2$ and $\Delta=0.4$. Landau levels are suppressed for $\Delta=0.4$ but a sharp peak around $E=0$ appears.} 
\label{4and25}
\end{figure}

Although Landau levels are observed in the low-energy DOS in the large $\Delta$ regime ($0.14\lesssim \Delta \lesssim 0.3$), for very large $\Delta$ ($\Delta \gtrsim 0.35$), the Landau levels disappear.  Fig. \ref{4and25}, shows a comparison of the low-energy DOS for $\Delta=0.2$ and $\Delta=0.4$ in the presence of a magnetic field $l=48$. The DOS for $\Delta = 0.4$ shows no Landau levels and a sharp peak at $E=0$ which can be attributed to the commensurability effect. This is further discussed in appendix \ref{a1}. 

As $\Delta$ is increased, large gaps appear within
the Fermi arcs.  This is seen in the spectral function, shown
in Fig. \ref{spall}(d) for $\Delta=0.4$ and zero magnetic field.
Assuming that the semiclassical process of
the formation of the Landau levels\cite{anderson} involves tunnelling of
particles (holes) across the gaps within the Fermi arcs, one expects
the magnetic breakdown phenomenon not to occur if the gaps are too
large. This may explain why the Landau levels are suppressed for very large
$\Delta$.

\begin{figure} [tbph] 
\includegraphics[scale=.21]{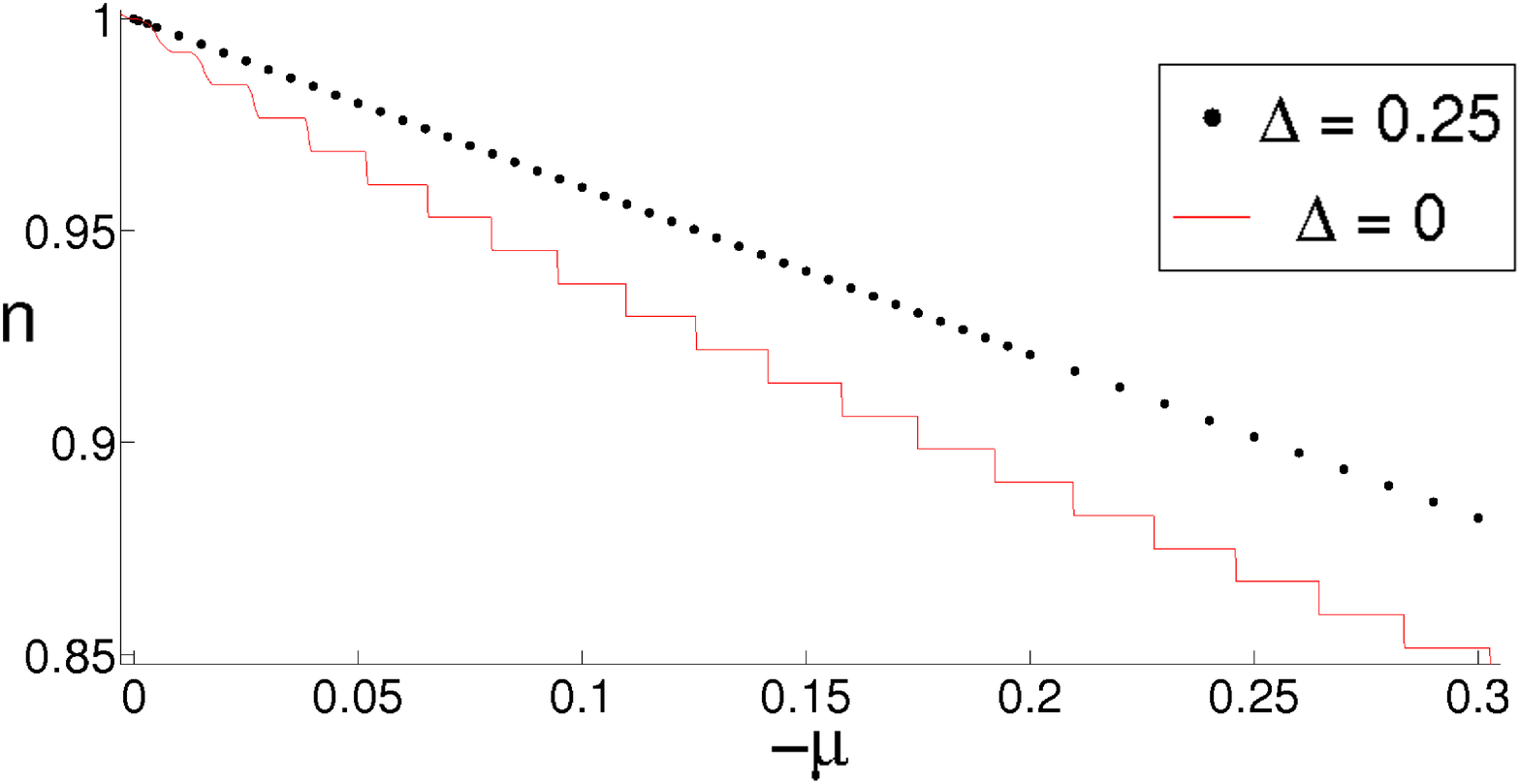}
\caption{Density of electrons versus $-\mu$ for the magnetic field of $l=16$ and two cases of $\Delta=0.25$ and $\Delta=0$. Unlike $\Delta=0$, the density does not exhibit a stepped behavior for $\Delta=0.25$.} 
\label{nvsm}
\end{figure}
Landau-type quantum oscillatory behavior has previously been discussed in the context of a particular model of a `Fermi-arc metal'.\cite{pereg-barnea} In that model, parts of the FS of a metal are artificially gapped out by restricting superconducting pairing to the antinodal regions of momentum space in order to get a FS that consists of Fermi arcs. The $\pi$-striped model, which is based on a specific microscopic mechanism and has no such restriction, differs from the Fermi-arc metal of Ref. 33 in that the one-electron spectral function has non-zero (but possibly very small) weight along continuous lines in $k$-space. 

At a more basic level, the behavior of a $\pi$-striped superconductor is strikingly different from that of a metal, in spite of the fact that both exhibit a Fermi surface.  In a metal, the particle density $n$ versus $\mu$ exhibits a stepped behavior in the presence of a constant magnetic field. In contrast, the particle density in a $\pi$-striped superconductor changes smoothly as a function of $\mu$ as shown in Fig. \ref{nvsm} for a large magnetic field of $l=16$. Furthermore, we find that the low-energy DOS behavior of the $\pi$-striped superconductor is rather insensitive to the change in $\mu$. In other words, no oscillatory behavior of the DOS at $E=0$ is observed as $\mu$ is varied except for finite size effects. This is in contrast to the result for the simple Fermi-arc metal model.\cite{pereg-barnea}  However,  quantum oscillations are induced by changing the magnetic field, not the chemical potential, and consequently could still be observed for a $\pi$-striped superconductor.

\section{Specific Heat}
In this section, we provide specific heat calculations in the absence and presence of a magnetic field to make connections to experiments on the
cuprates. In advance, we note that the $\sqrt{B}$
dependence of the Sommerfeld coefficient, $\gamma$, in the
cuprates is not present in the $\pi$-striped superconducting model as there is a finite DOS at $E=0$. However, it will be shown that some features of the specific heat in the cuprates are consistent with this model. One can calculate the specific heat by using the relationship $c=T\frac{\partial S}{\partial T}$. For a system of quasiparticles, the entropy is given by\cite{degennes}
\begin{equation}\label{entropy}
S=-k_{B}\sum_{p \alpha}[f_{p}\ln f_{p}+(1-f_{p})\ln (1-f_{p})]
\end{equation}
where $\alpha$ is the spin state and $f_{p}$ is the Fermi-Dirac distribution function given by
\begin{equation}
f_{p}=\frac{1}{1+\exp(\epsilon_{p}/k_{B}T)}
\end{equation}
and $\epsilon_{p}$ is the energy of the quasiparticle associated with the state $p$. Converting the sum over states in Eq. \ref{entropy} to an integral over energy brings in the DOS. It should be noted that, in this study, the DOS is not calculated self-consistently as there is no microscopic Hamiltonian defined.  Furthermore, it is assumed that the magnitude of the pairing interaction is fairly constant at low temperatures so that the quasiparticle spectrum is unchanged as temperature increases.

\begin{figure} [tbph] 
\includegraphics[scale=.21]{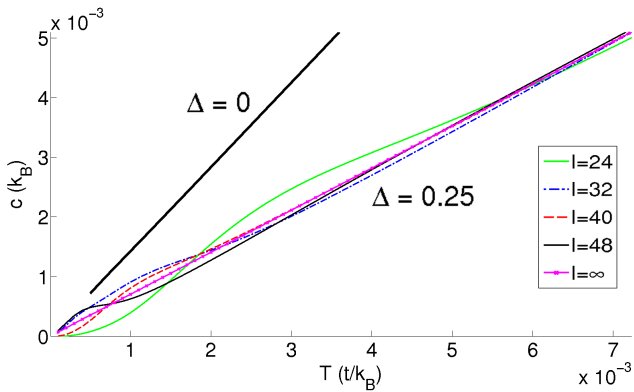}
\caption{Specific heat in the presence of various fields as a function of temperature for $\Delta=0.25$ and $\mu=-0.3$. The behavior of the curves at very low temperatures is significantly affected by the commensurability effect. The heavy line shows the specific heat in zero field for $\Delta=0$ and $\mu=-0.225$ corresponding to $1/8$ doping. The slope associated with the linear behavior is about two times that of the slope for $\Delta=0.25$ in zero field (noted by $l=\infty$).} 
\label{cB}
\end{figure}

The specific heat at zero field for a $\pi$-striped superconductor as a function of temperature is shown in Fig. \ref{cB} for $\Delta=0.25$ at $1/8$ doping. The slope associated with the linear behavior is $\gamma=0.70 k_{B}^{2}t^{-1}$ per site and is directly proportional to the DOS at $E=0$, which is $0.21t^{-1}$ per site. The slope is about half of that of $\Delta=0$ at $1/8$ doping. The low-energy specific heat of various fields as a function of temperature is also shown in Fig. \ref{cB} for $\Delta=0.25$ and $\mu=-0.3$. As expected, all the curves converge to that of zero field as the temperature increases. However, at very low temperatures, the specific heat behavior for different fields is significantly affected by the commensurability effect. This is seen in the nearly zero slope of the curves for odd $m$ (recall $l=8m$) as $T\rightarrow 0$.

Low temperature electronic specific heat measurements of a
cuprate\cite{riggs} indicate a relatively large DOS at $E=0$ which can not
be explained by the presence of disorder in a d-wave superconductor.
Taking the lattice constant of a typical cuprate to be $a=3.85\AA$, the
specific heat effective mass becomes $m^{*}/m=0.34 eV/t$. A rather wide range of values
has been used for $t$.\cite{shirit,yao} Within a simplified
nearest-neighbor hopping only model as used here, we obtain $m^{*}/m=1.36$ for $t
\approx 0.25eV$.\cite{goswami} This value corresponds
to $\gamma\approx 1.98$ mJ$\cdot$ K$^{-2}\cdot$mol$^{-1}$ which is
consistent with the specific heat measurements for cuprates in the absence
of a magnetic field, $\gamma\approx 1.85$ mJ$\cdot$
K$^{-2}\cdot$mol$^{-1}$.\cite{riggs}

Riggs et al. have studied the low temperature specific heat as a function of magnetic field up to very high fields (50
T) and observed quantum oscillations.\cite{riggs} This allowed them both to measure the magnitude of the specific heat
in what is presumably the normal state, and also to determine the cyclotron
effective mass associated with the quantum oscillations. Then if one assumes the normal state has broken translation symmetry, modeling the arrangement of electron and hole pockets in the
Brillouin zone and using the measured cyclotron effective mass, one can estimate what the specific heat should be. The result
is much larger than the specific heat that they observed.\cite{riggs,zlatko}  The same problem was also noted in a
theoretical study\cite{yao} based on FS reconstruction where the calculated specific heat was larger than the measured value.  

In our calculations for a $\pi$-striped superconductor, it was found that, even though there exist several FS pockets in the
BZ, a single set of  Landau levels is observed above and below the Fermi energy.  The relation between the slope of Fig.
\ref{TvsB}, defining the cyclotron effective mass obtained from the spacing of Landau levels, and the DOS at $E=0$ is the
same as for the $\Delta=0$ case. This implies that, as for the $\Delta=0$ case, the cyclotron effective mass, $m_{c}$, is
equal to the specific heat effective mass, $m^{*}$, for large values of $\Delta$. Consequently, the quantum oscillations
in the specific heat and the magnitude of the specific heat which is observed in Ref.~\onlinecite{riggs} could be
consistent with the behaviour of a $\pi$-striped superconductor state induced by large magnetic fields, rather than a
striped metallic state with no pairing gap as is often assumed. However, as noted earlier, the ideal $\pi$-striped model
(with no uniform $d$-wave component) is not expected to give a $\sqrt{B}$ background, which also appears to be a feature
of the experiments.\cite{riggs} In addition, since we cannot study small changes in $B$ we make no prediction about the
spacing of the observed quantum oscillations.
\section{Discussion and Conclusion}
In this work we have studied the requirements for having quantum oscillations in a model of a $\pi$-striped superconductor. For a large range of values of the pairing interaction, the FS corresponds to closed loops while the one-particle spectral function exhibits Fermi arcs in k-space. Our main finding  is that Landau levels are seen in the low-energy DOS of the $\pi$-striped superconductor in a large range of the magnetic field, which indicates the possibility of quantum oscillations.  We find that low-energy Landau level formation persists even though particle and hole levels are mixed by the pairing interaction.  Other theoretical studies of quantum oscillations in the cuprates are typically based on FS reconstruction of a metallic state and involve multiple pockets and frequencies.\cite{millis,ddensity} Furthermore, the pockets associated with those studies are located where the ARPES experiment shows a large pseudogap.  By contrast, the $\pi$-striped superconductor exhibits a unique low-energy Landau level set that is only due to the Fermi arc part of the spectral weight function.  

Since our numerical studies are restricted to satisfying $l=8m$, we cannot change the magnitude of the magnetic field continuously or in small steps. In addition, the Landau levels are located symmetrically around $E=0$ for the discrete values of the magnetic field that we can study. As a result it is not possible, from these calculations, to find the FS area associated with quantum oscillations that would be observed by conventional experimental methods.\cite{onsager} However, we can make conjectures about FS areas that might be observed, based on our analysis. We expect that any semiclassical trajectory describing the formation of Landau Levels should have the following characteristics: 1) The trajectory should use all parts of the FS.  2) Andreev scattering needs to occur at least at one point during the Larmor precession because Landau levels are a coherent mixture of particles and holes.  3) Magnetic breakdown is likely involved in Landau level formation because once the gaps within the Fermi arcs become too large, the Landau levels disappear.

It is also useful to compare the behaviour of the $\pi$-striped superconductor to the Fermi-arc metal, in which a new mechanism for quantum oscillations is proposed that is not based on FS reconstruction.\cite{pereg-barnea} For that model, it was shown, based on a semiclassical approach, that the frequency of quantum oscillations is proportional to the Fermi arc length. In the $\pi$-striped superconductor, the fact that a quasiparticle with wave vector $k$ is coupled to ones with wave vectors $-k-q_{x}$ and $-k+q_{x}$ provides a different scattering mechanism which changes the semiclassical motion of quasiparticles. Consequently, the semiclassical trajectories of the two studies are expected to be different. Although Landau level formation in the Fermi-arc metal resembles what we have seen in the $\pi$-striped model, the DOS at $E=0$ shows an oscillatory behavior as a function of $\mu$ for the Fermi-arc metal which is inconsistent with our study. 

We note that the observation of quantum oscillations corresponding to small Fermi surface pockets supports the scenario of translational symmetry breaking and Fermi surface reconstruction, whether due to charge or spin density waves or to modulation of the d-wave gap, as discussed in this paper. Indeed recent experimental results support the connection between stripe formation and quantum oscillations.\cite{lali} Our calculations show that a modulated d-wave superconductor can support Landau levels and quantum oscillations but we are unable to make detailed comparisons to quantum oscillation experiments because of the restriction to commensurate vortex lattices. One might expect there to be observable differences between quantum oscillations in the presence of charge or spin stripes and superconducting stripes, due to the Andreev reflection and particle-hole mixing involved in the formation of Landau levels in the latter case. Therefore, it would be of interest to study modulated superconductivity within a framework which allows the magnetic field to be varied continuously to more directly connect to the quantum oscillation experiments on the cuprates. Possible approaches would be to use random vortex lattices, as was done by Chen and Lee,\cite{lee} or to develop a semiclassical approximation that allows magnetic unit cells of arbitrary aspect ratios.
\begin{acknowledgements}
The authors acknowledge useful discussions with Marcel Franz and Steven Kivelson. This work was supported by the Natural Sciences and Engineering Research Council of Canada and the Canadian Institute for Advanced Research.
\end{acknowledgements}
\appendix
\section{Commensurability effects} \label{a1}
In this appendix, we discuss the commensurability effect mentioned in Sec. \ref{section:dos}. We found that the DOS at $E=0$ exhibits a periodic behavior as a function of $l$ or $1/\sqrt{B}$ for large values of $\Delta$. If $m$ is even, the two vortices in a magnetic unit cell are in perfectly equivalent positions with respect to the spatially modulated gap and the DOS at $E=0$ is nonzero. In contrast, for odd $m$, the gap on the right of one vortex is positive but on the right of the other vortex is negative as can be seen in Fig. \ref{config}. The DOS at $E=0$ is zero for odd $m$.

The same kind of commensurability effect is also seen in a uniform d-wave superconductor.\cite{melikyan} There, due to strong internodal scattering, the DOS at zero energy exhibits a periodic behavior as a function of $k_{d}l$ where $k_{d}$ is the $k$-space half distance between the nearest nodes of the d-wave superconductor. Specifically, depending on whether $n$ is odd or even in $k_{d}l=\pi n$, the DOS around $E=0$ shows a linear or gapped behavior. For a $\pi$-striped superconductor, the relevant $k$-space half distance is $\pi/8$ which leads to a periodicity of $\delta l=16$ for the DOS at zero energy as a function of $l$. So the commensurability effect seen in a $\pi$-striped superconductor is most likely due to interference effects. This suggests that the nonzero DOS at $E=0$ for even $m$ is due to constructive interference of particle and hole waves, while the gapped behavior for odd $m$ is due to destructive interference.
\begin{figure} [tbph] 
\includegraphics[scale=.3]{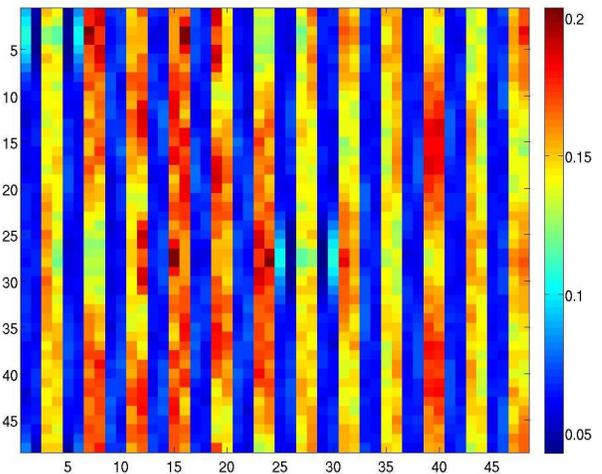}
\caption{Local density of electrons due to the low-energy states within 0.001t of $E=0$ for $l=48$ and $\Delta =0.4$.} 
\label{doe2}
\end{figure} 

For very large $\Delta$, a sharp peak develops near $E=0$ for even $m$ only, as shown in Fig. \ref{4and25}. It appears that the origin of the peak can be traced back to the non-zero DOS at $E=0$ for smaller $\Delta$, and consequently is related to the commensurability effect. The fact that the low-energy  Landau levels disappear when the peak at $E=0$ is sharp suggests that the commensurability effect is competing with the Landau level formation. Fig. \ref{doe2} shows the real space representation of the the states under the sharp peak at $E=0$ where a pattern of stripes of low and high particle density is clearly visible. On average, the ratio of the density of the higher density stripes to the lower density is 2.35 for $\Delta =0.4$, which is larger than the value 0.2 in the absence of a magnetic field (see appendix \ref{a2}). It is not clear whether these effects are due to the existence of a strong commensurability effect or  due to the presence of large gaps within the Fermi arcs.
 
\section{Periodic Andreev state}\label{a2}
\begin{figure} [tbph] 
\includegraphics[scale=.38]{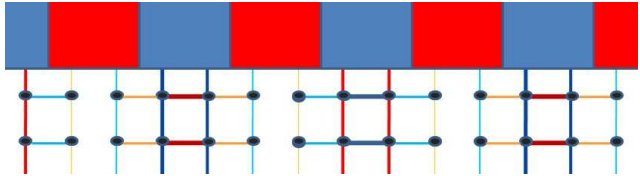}
\caption{High (red stripe) and low (blue stripe) density structure of the low-energy particles relative to the modulated d-wave gap. High (low) density is indicated by a red (blue) stripe.} 
\label{stripe}
\end{figure}   
In this appendix, a type of Andreev state that is seen for the low-energy particles (holes) in the absence of a magnetic field and persists in the presence of a magnetic field is discussed. We have already examined how the spectral weight of the low-energy states of a $\pi$-striped superconductor changes as the pairing amplitude $\Delta$ is varied. We can also look at the real space representation of these states. Our main finding is that, for large enough values of $\Delta$ where the shape of Fermi arc is assumed in the spectral function (see Fig. \ref{spall}), the real space representation of the low-energy states exhibits a periodic stripe structure with the periodicity of four lattice sites. The stripe structure corresponds to higher and lower density of low-energy electrons and holes. Each stripe has a width of two lattice constants and the higher density stripes are located exactly where the order parameter is minimum as shown in Fig. \ref{stripe}. The density ratio of the higher density stripes to the lower density ones increases as $\Delta$ increases. The ratio is approximately 1.5 for $\Delta=0.2$ and 2 for $\Delta=0.4$. 

This stripe structure occurs not only at the Fermi energy, but also at energies near Fermi energy. The origin of these stripes is simple. They are formed due to the constructive interference of the low-energy electron (hole) waves with their doubly Andreev scattered counterparts. Since the probability of being Andreev scattered twice increases with $\Delta$, the density difference between higher and lower density stripes increases accordingly. One can think of the stripe structure as a periodic Andreev state. For a wave vector at the end-points of the Fermi arcs, the doubly scattered wave vector is also located at the end-point of another Fermi arc. This means that the weights of both interfering waves are large and as a result they contribute considerably to the formation of the stripe structure. Moving toward the center of the Fermi arc, the coupling interaction decreases and the periodic Andreev structure is less likely to be formed. This is why, as mentioned in Ref. 24, the stripe structure is mainly due to states near the end-points of the Fermi arcs.
\bibliography{bib}
\bibliographystyle{unsrt} 
\end{document}